\documentclass[preprint]{aastex}

\newcommand{\beq}{\begin{equation}}
\newcommand{\eeq}{\end{equation}}

\shorttitle{X-Ray Pulsar Cyclotron Line Variability}
\shortauthors{C. Weth}

\begin{document}

\title{A Doppler Shift model to explain the Cyclotron Line Variability
       in X-Ray Pulsars} 
\author{C. Weth}
\affil{Theoretische Astrophysik T\"ubingen} 
\affil{Auf der Morgenstelle 10, 72076 T\"ubingen}
\email{weth@tat.physik.uni-tuebingen.de}

\keywords{line: formation --- relativity --- stars: magnetic fields
--- stars: neutron --- X-rays: binaries }

\begin{abstract}
A simple model to explain the phase-dependence of the
cyclotron absorption line observed in many X-Ray pulsars is
presented. It includes several relativistic effects, namely
gravitational redshift, gravitational light deflection, and -- most
important -- doppler shift of the photon energy in the infalling
plasma. It is shown that previous estimates of neutron star magnetic
fields neglecting the last effect give results which are too small by
about a factor of two. Finally the developed equations are used to
determine the neutron star radius and magnetic field of Her X-1, but
the large uncertainties in the geometry prevent the results from
seriously constraining the parameters.
\end{abstract}

\maketitle

\section{Introduction}

Cyclotron absorption lines are a important diagnostic tool in the
analysis of X-Ray pulsars since they provide a way to measure the
magnetic field of the neutron star directly. First discovered in Her
X-1 \citep{truemper78}, there are now about a dozen systems in which
they can be seen \citep{makishima99}. 

In the last 20 years, phase-resolved spectroscopy of the cyclotron line
region has become possible and has been done for several systems
(Her X-1: \citet{voges82,soong90}, 4U 1538-52: \citet{clark90},
4U 1907+09, Vela X-1: \citet{makishima99}, Cen X-3:
\citet{burderi00}). In most of them the cyclotron line energy varies
with the pulse phase within 10 to 20\%.

Model calculations of radiative transfer in slabs and columns by
several authors yielded viewing-angle dependent cyclotron line
features, e. g. \citet{meszaros85a,isenberg98}.  However, the line
variations obtained there are not sufficient to explain the observed
values. \citet{brainerd91} find that absorption in the infalling
plasma of an accretion column yields a much stronger feature at lower
energy than the one obtained in the slab models, the energy of which
is angle-dependent due to the bulk motion of the infalling plasma. A
second consequence of the high velocity (the free-fall velocity is
about $0.5c$ at the neutron star surface!) is that most of the
scattered photons are directed back onto the neutron star, so the line
is seen in absorption.

In the following section, a simple analytic model that gives an
analytic formula for the cyclotron line energy seen by the observer
will be introduced. After discussing some of its properties, the
model is applied to Her X-1 where the geometrical parameters have been
analyzed \citep{blum00}. Finally, the results and implications are
discussed.


\section{A simple model}

Consider a neutron star with radius $R_{\mathrm{N}}$ and gravitational
(Schwarzschild) radius $R_{\mathrm{S}}$ in the Schwarzschild metric. Assume that
the region around the magnetic poles where the accreted plasma hits
the neutron star is small, so that $B\approx const$ and the direction of the
infalling matter is orthogonal to the neutron star surface;
the polar cap itself is assumed to be plane. Quantities in the Lorentz
system resting on the neutron star surface are marked with a hat in
the following, whereas the ones in the rest system of the infalling
plasma have an index 0.

Neglecting radiation pressure, the local free-fall velocity on the
surface is $\hat \beta_{\mathrm{ff}}=\hat v_{\mathrm{ff}}/c =
\sqrt{R_{\mathrm{S}}/R_{\mathrm{N}}}$. If the matter is decelerated
above the surface, a factor $\eta \le 1$ can be introduced so that
$\hat \beta =\eta \hat\beta_{\mathrm{ff}}$.

In the nonrelativistic limit ($B \ll B_{\mathrm{crit}} \approx 4.4
\times 10^{9} \mathrm{T}$), the cyclotron energy in the frame at the
neutron star surface is

\beq
\hat E_{c} = \hbar\omega_c = \frac{\hbar e B}{m_e} = 11.58~B_{8}~\mbox{keV}
\eeq

where $B_8$ is the magnetic field in $10^8$ Tesla.

In addition to the magnetic field strength at the neutron star
surface, there are several effects that determine at which energy the
cyclotron absorption line in the spectrum is seen by a distant
observer: gravitational redshift, gravitational light bending, and
the doppler effect due to the bulk motion of the infalling
plasma.

The first one of them, gravitational redshift, simply shifts the
spectrum to a lower energy by a factor of $\sqrt{1-R_{\mathrm{S}}/R_{\mathrm{N}}}$ (and
decreases the intensity, which is of no interest here).
However, the radius of the neutron star is generally not known, so
that results of theoretical works (e. g. \citet{shapiro,meszaros87,wu91}),
models fits \citep{bulik95} or other indirect measurements
\citep{zhang97} have to be used. Typical values range from $2$ to
about 4 $R_{\mathrm{S}}$.

Gravitational light deflection changes the direction under which
radiation emerging from the hot spot is seen. Again, the strength of
this effect depends on the neutron star radius. Its effect can be
described by an elliptic integral (see e.g. \citet{weinberg}):

\beq
\Delta\theta = \int\limits_{R_{\mathrm{N}}}^\infty \frac{b}{r^2}
\frac{dr}{\sqrt{1-A(r)b^2/r^2}},
\eeq

where $A(r)=1-R_{\mathrm{S}}/r$ and $b$ is the impact parameter of the photon,
which is given by $b=R_{\mathrm{N}}
\sin\hat\vartheta/\sqrt{1-R_{\mathrm{S}}/R_{\mathrm{N}}}$. $\hat\vartheta$ is the
propagation angle of the photon with respect to
the radial direction at the surface.

The third effect, doppler shift, is responsible for the angle- (and
thus phase-) dependence of the line energy: The resonance occurs (for
cold plasma) when the photon energy equals the cyclotron energy
\textit{in the plasma rest frame}. While the magnetic field strength
is not altered by the Lorentz boost as long as the motion of the
matter is along the field lines, the photon energy is:

\beq
E_0 = \hat E \frac{(1+\cos\hat\vartheta \hat\beta)}{\sqrt{1-\hat\beta^2}}
\eeq

In summary, the spectrum of the radiation going in the direction
$\hat\vartheta$ at the neutron star surface will be seen by an observer
at $\hat\vartheta+\Delta\theta$; the absorption line will appear at

\beq
\label{ecobs}
E_{c,\mathrm{obs}}(\theta_{\mathrm{obs}}) = \frac{\hbar e B}{m_e}
  \frac{\sqrt{1-(\eta\hat\beta_{\mathrm{ff}})^2}}{(1+\cos\hat\vartheta\,\eta\hat\beta_{\mathrm{ff}})}
  \sqrt{1-R_{\mathrm{S}}/R_{\mathrm{N}}}
  \stackrel{\eta=1}{=}
  \frac{\hbar e B}{m_e}
  \frac{1-R_{\mathrm{S}}/R_{\mathrm{N}}}{(1+\cos\hat\vartheta \sqrt{R_{\mathrm{S}}/R_{\mathrm{N}}})}
\eeq

\begin{figure}
\plotone{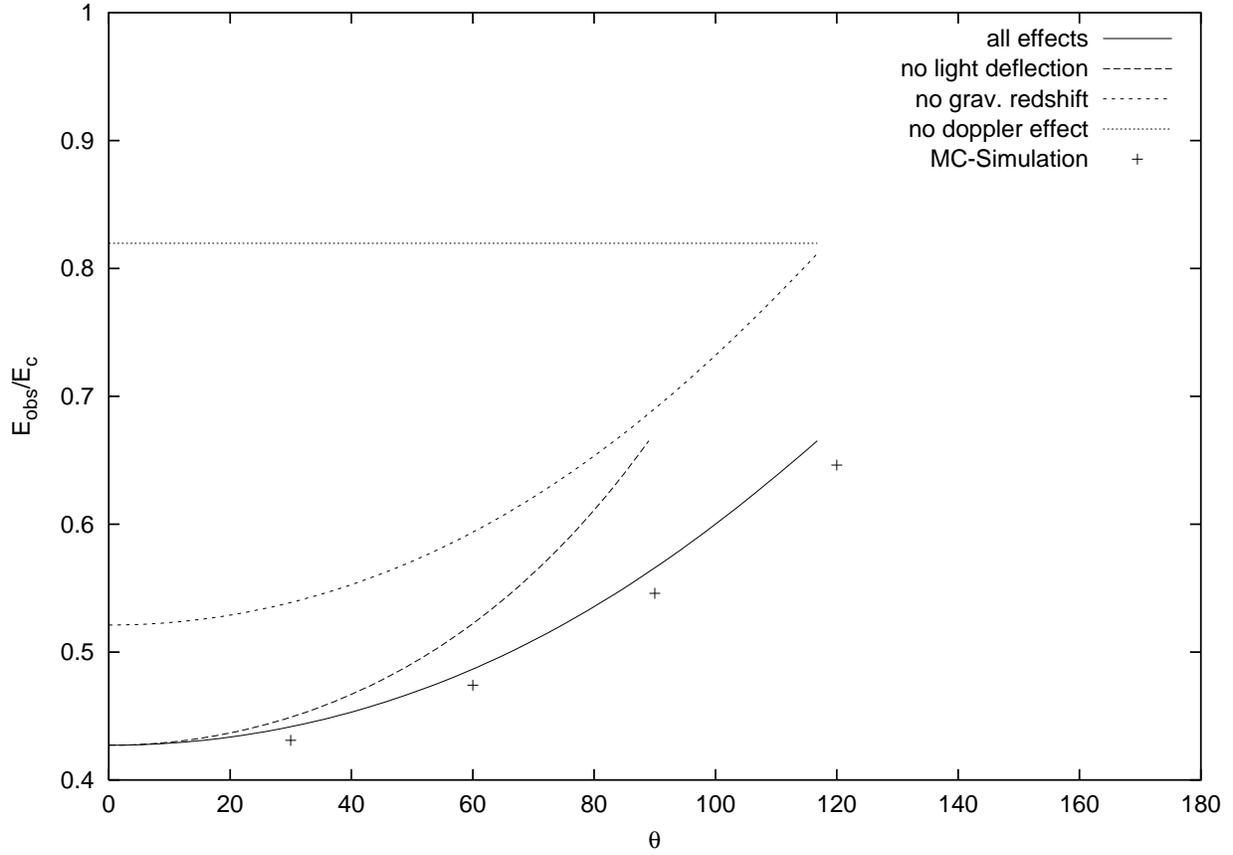}
\caption{Cyclotron line energy seen by the observer in units of the
cyclotron energy on the neutron star surface (solid line). The other
curves demonstrate how neglecting one of the relativistic effects
changes the result.  The crosses are the result of a
Monte-Carlo-Simulation of a more sophisticated model. The neutron star
radius is $3R_{\mathrm{S}}$; no radiation pressure is included.
\label{energies}}
\end{figure}

Figure \ref{energies} shows the measured cyclotron line energy in units of
the line energy in the plasma rest frame for a neutron star of
$R_{\mathrm{N}}/R_{\mathrm{S}}=3$. Note that for this ratio, the line stops at the angle
$\theta_{\mathrm{obs}}\approx 120^\circ$, because photons would have to pass
through the neutron star in order to get there. However, in more
sophisticated models, photons can be scattered at some point above the
hot spot and reach these regions.

%

Several more curves are shown in the figure: The constant one is the
line one gets neglecting the doppler effect, which has been done in
most publications computing the magnetic field in X-Ray binaries so
far. The dotted lines are without gravitational redshift or
gravitational light deflection; the crosses are the results of a
Monte-Carlo-Simulation that uses a realistic geometry and
nonrelativistic cold cross sections. Details of this model will be
published elsewhere (Weth et al., in preparation). These line energies
are systematically lower than the ones of the simple model; the reason
for that is that scattering takes place also above the hot spot, where
the magnetic field drops steeper than $r^{-3}$
\citep{wasserman83}. The difference in the plot corresponds to a
height of about 150m.

In real systems, there are two accreting polar caps, so what the
observer sees is the sum of two spectra at different angles, which are
$\theta_{1}$ and $\theta_2=180^\circ-\theta_1$ in the simplest
case. Because the luminosity depends on the direction, too, at some
angles the line feature of one cap will be hidden in the background of
the other one. In most cases however, these two lines will be too
close to each other to be resolved.


\section{Application to Her X-1}

A quantity which can be measured is 

\beq
\xi(R_{\mathrm{N}}/R_{\mathrm{S}}) = \frac{E_{c, \mathrm{max}} }{ E_{c,\mathrm{min}}}
\eeq

Its value is always smaller than the theoretical maximum
which can be obtained by taking the cyclotron energies seen at
$\theta_{\mathrm{obs}}=0$ and $\theta_{\mathrm{obs}}=180^\circ$. It depends only on the
neutron star radius and on $\eta$. However, this does not
constrain the parameter $R_{\mathrm{N}}/R_{\mathrm{S}}$ significantly unless $\xi \ge 2$,
which has not been found yet in any system.

In the case of Her X-1 the geometry (i. e. inclination of the rotation
axis and position of the magnetic poles) is known \citep{blum00}, so a
model value of $\xi$ can be computed by using the appropriate
angles. Unfortunately Her X-1 is not ideal for our purpose because of
its high luminisity of $L_{37}=2.0$ \citep{nagase89}, so that
radiation pressure cannot be neglected here. In consequence, there are
three unknown parameters ($R_{\mathrm{N}}$, $B$, and $\eta$), but only
two quantities that can be measured ($E_{c, \mathrm{max}}$ and
$E_{c,\mathrm{min}}$).

In order to compute the parameters, an additional assumption is
needed. Two approaches will be tried: First
$R_{\mathrm{N}}/R_{\mathrm{S}}$ is fixed to a specific value, in the
second one radiation pressure is completely neglected by setting
$\eta=1$.

Adopting $E_{\mathrm{max}}=36.9\pm0.3$ keV, $E_{\mathrm{min}}=30.3\pm1.0$ keV
\citep{soong90} and $i=83^\circ\pm4^\circ$ and
$\theta_{\mathrm{pole}}=20^\circ\pm10^{\circ}$ \citep{blum00}, and further
assuming a neutron star radius of $R_{\mathrm{N}}=3.0 R_{\mathrm{S}}$, equation
(\ref{ecobs}) yields $\eta=0.9\pm0.5$, which is consistent with no
radiation pressure. Most of the error is due to the large uncertainty
in $\theta_{\mathrm{pole}}$, the positions of the magnetic poles relative to
the rotation axis.

\begin{figure}
\plotone{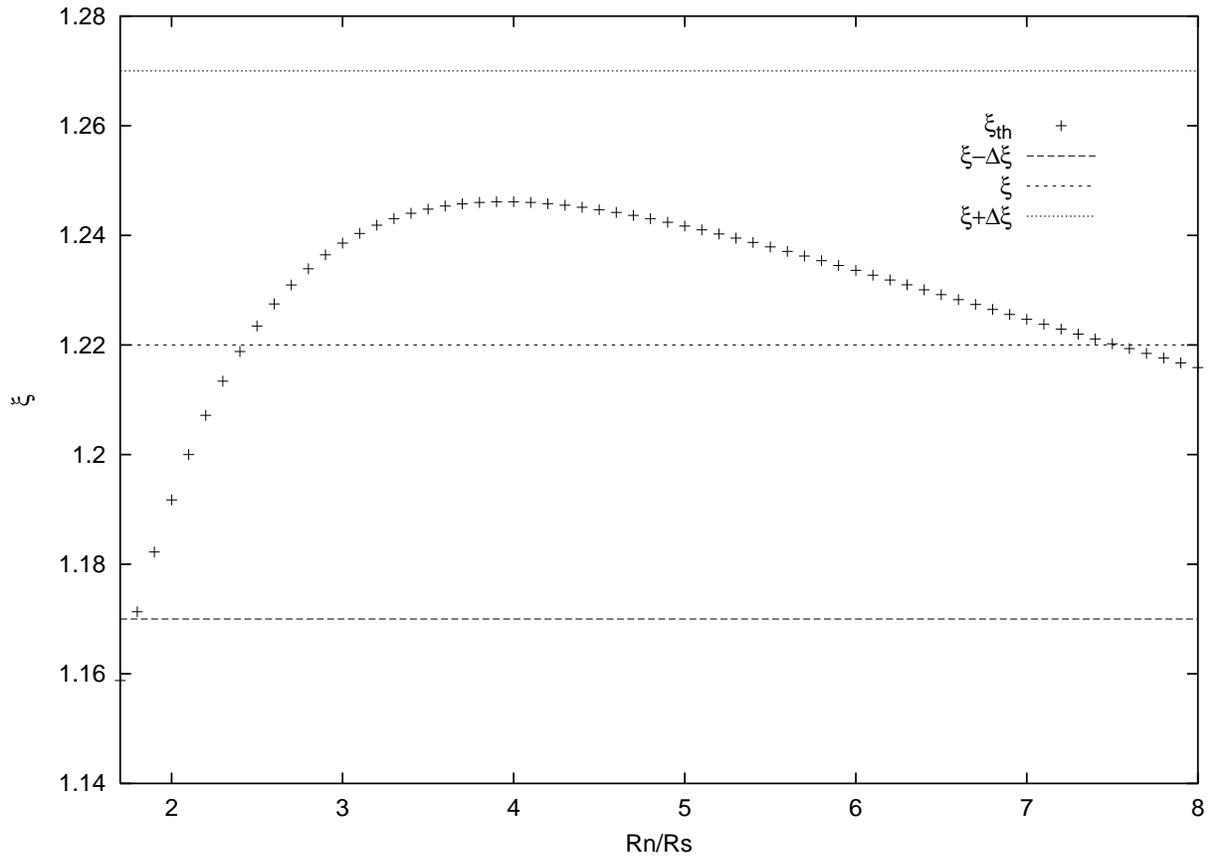}
\caption{$\xi(R_{\mathrm{N}}/R_{\mathrm{S}})$ without radiation pressure. The horizontal
lines are the measured $\xi$ and the error range.}
\label{xi}
\end{figure}

Fixing $\eta=1$ and determining the neutron star radius gives the
right range ($2.4 \le R_{\mathrm{N}}/R_{\mathrm{S}} \le 7.5$), but including the error in
$\xi$ causes the upper limit to be shifted beyond $12 R_{\mathrm{S}}$
(Fig. \ref{xi}).

%

Finally, the magnetic field at the neutron star surface is computed
from the previous results ($\eta=0.9\pm0.5$, $R_{\mathrm{N}}=3.0
R_{\mathrm{S}}$ fixed). This yields $B_8=5.0\pm0.9$, the main cause
for the large error being the uncertainty in $\eta$. Varying
$R_{\mathrm{N}}$ changes the result by about $\pm 1$. This value of
$B$ is considerably higher than the one given by \citet{voges82},
which was $3.4$.


\section{Discussion}

With the doppler shift in the accretion column, a simple explanation
for the phenomenon of the phase-dependent cyclotron line energy was
found; it should occur in almost every system except those in which
the radiative pressure is so high that the velocity of the infalling
matter becomes small compared to $c$. However, there are some
problems: Cyclotron line absorption might not only take place close to
the hot spot, but also higher in the accretion column, especially when
the column is optically thick; this shifts the cyclotron line centers
to lower energies. In this case, not the neutron star radius
$R_{\mathrm{N}}$ is obtained, but the distance of the emission region
from the neutron star center (which is at least an upper limit to the
radius). Another problem is that not the full range of cyclotron line
energies might be seen - if the line feature from one cap falls into a
region with high continuum from the other one, it will be hard to
detect.

Still, the range of values obtained for the neutron star radius is where
one expects the actual radius to be. The resulting magnetic field
strength is significantly higher than the usual estimates ($B_8\approx
5$ instead of 3), which only take into account gravitational redshift
but not the doppler effect due to the moving plasma. This fact was
already mentioned by \citet{brainerd91}.

A prediction of this simple model is the presence of two absorption
lines (one from each emission region) in the spectrum that move
anticyclically with the pulse phase. This should make them
distinguishable from a second harmonic, the energy shift of which is
in phase with the one of the fundamental line. The detection of
absorption features from both hot spots will only be possible for
systems where the angle between magnetic and rotation axis is large,
so that one can see the emission regions at all angles; in other
systems, they will be too close to each other to be resolved.

On the other hand, for not too large neutron star radii, the
absorption feature from the surface and the one from the column should
be distinct \citep{brainerd91}. However, it is not clear if the
surface feature is visible at all, since it might be filled up by
photon spawning effects \citep{isenberg98}. The column absorption
feature is probably not changed too much by those, because most
scattered photons are advected back onto the neutron star.


\acknowledgments

I am grateful to Ute Kraus for valuable discussions. This work was
supported by the DFG (Deutsche Forschungsgemeinschaft).


\bibliography{MyBibliography}
\bibliographystyle{apj}

\end{document}